\documentclass{emulateapj}

\usepackage{xspace}
\usepackage{graphicx, floatflt}

\newcommand{\NH}{\mbox{${\rm N}_{\rm H}$}}        % Defines NH
\newcommand{\etal}{{\it et al.}\xspace}
\newcommand{\chandra}{{\sl Chandra}\xspace}

\begin{document}

\title{X-ray Dust Scattering at Small Angles: The Complete Halo around
  GX13+1} 

\author{Randall K. Smith} \affil{NASA Goddard Space Flight
  Center, Greenbelt, MD 20771} \affil{Department of Physics and
  Astronomy, The Johns Hopkins University, 3701 San Martin Drive,
  Baltimore, MD 21218}

\begin{abstract}

The exquisite angular resolution available with \chandra should allow
precision measurements of faint diffuse emission surrounding bright
sources, such as the X-ray scattering halos created by interstellar
dust.  However, the ACIS CCDs suffer from pileup when observing bright
sources, and this creates difficulties when trying to extract the
scattered halo near the source.  The initial study of the X-ray halo
around GX13+1 using only the ACIS-I detector done by \citet{SES02}
suffered from a lack of sensitivity within $50''$\ of the source,
limiting what conclusions could be drawn.

To address this problem, observations of GX13+1 were obtained with the
\chandra HRC-I and simultaneously with the RXTE PCA.  Combined with
the existing ACIS-I data, this allowed measurements of the X-ray halo
between 2-1000''.  After considering a range of dust models, each
assumed to be smoothly distributed with or without a dense cloud along
the line of sight, the results show that there is no evidence in this
data for a dense cloud near the source, as suggested by
\citet{Xiang05}.  Finally, although no model leads to formally
acceptable results, the \citet{WD01} and nearly all of the composite
grain models from \citet{ZDA04} give poor fits.
\end{abstract}

\keywords{dust, extinction --- scattering --- X-rays: ISM}
\section{Introduction} 

Practically every band of the electromagnetic spectrum affects or is
affected by interstellar (IS) dust grains.  In the IR, PAHs emit lines
and small grains emit continuum radiation; in the UV/optical, small
grains both extinct and scatter light.  In X-rays, large dust grains
($>0.1 \mu$m) scatter X-rays, creating halos around point sources.
The classic paper by \citet[][MRN77]{MRN} used the observed optical
extinction to determine the size distribution of dust grains between
0.005-0.25$\mu$m.  Newer models, such as \citet[][WD01]{WD01}, have
extended the modeling to include polycyclic aromatic hydrocarbons
(PAHs) to match the observed IR emission as well as other constraints
on grain abundances.  Recently, \citet[][ZDA04]{ZDA04} found that a
wide range of dust compositions and size distributions could fit the
existing data, and suggested that new observational constraints from
X-ray halos are needed to select amongst these models.

X-ray dust scattering halos are created by the small-angle scattering
of X-rays as they pass through dust grains.  When an incoming X-ray
interacts with the electrons in a grain large compared to the X-ray
wavelength, the resulting Rayleigh scattering adds coherently in the
forward direction leading to small-angle scattering; see \citet{vdH57}
and \citet{ML91} for details, and \citet{Draine03} for a comprehensive
review.  More generally, the scattering problem can be posed as that
of a wave interacting with a sphere, in which case the Mie solution
applies \citep[{\protect{\it e.g.}}][]{SD98}.  In either approach, the
scattering depends largely on the grain size distribution, with lesser
dependencies on the grain composition and position along the line of
sight.

Observations of X-ray scattered halos have just begun to significantly
impact dust models.  \citet[][(SES02)]{SES02} described \chandra
observations of GX13+1 with the ACIS-I detector and showed that dust
grains do not have large ($\gtrsim 0.8$) vacuum fractions considered
by \citet{MW89}.  SES02 also found that the extremely large grains
found by {\sl Ulysses}\ in the solar neighborhood \citep{Landgraf00,
WSD01} do not seem to be common throughout the Galaxy.  Despite these
successes, SES02 could not distinguish between the MRN77 and WD01
models.  This was in part due to calibration uncertainties as well as
inherent limitations of the data.  Despite \chandra's excellent
angular resolution, ACIS-I observations of GX13+1 could not measure
the halo within $50''$, due to massive pileup in the ACIS-I detectors.
\citet{Draine03} and \citet{Xiang05} have both noted that this result
is therefore insensitive to dust near the source, as scattering from
dust within the last 25\% of the distance would lead to features
primarily within the excluded $50''$.  To address this shortcoming, I
obtained a short \chandra HRC-I observation of GX13+1.  The
multichannel plate design of the HRC-I is far less sensitive to large
count rates, which allows GX13+1's radial profile to be measured to
within 2'' of the source, far closer than previously possible.

%*********************************************************************

\section{Observations}

GX13+1 was observed simultaneously with the \chandra HRC-I and RXTE
Proportional Counter Array (PCA) on February 8, 2005 for 9.1 ksec
(ObsID 6093) and 6.2 ksec (P90173), respectively.  CIAO v3.3 software
was used to process the \chandra data, which showed significant
background flares in addition to the flux from the bright source.
Standard processing was used for the RXTE data.

\subsection{Selecting Good Events}

\begin{figure}
\includegraphics[totalheight=2.3in]{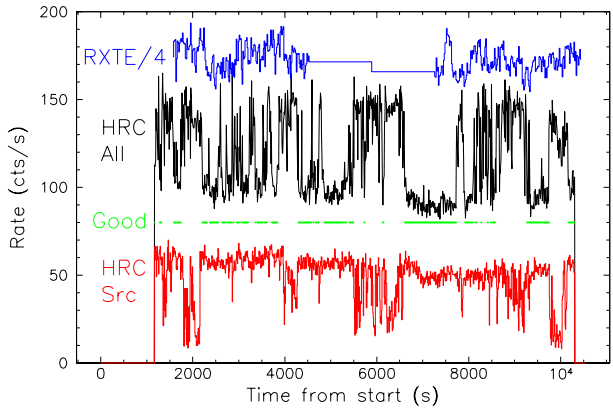}
\caption{GX13+1 lightcurve from HRC-I (red) and RXTE (in blue; divided
  by 4 for clarity).  The full HRC-I field is shown in black, and the
  selected ``good time intervals'' are shown in green.  When the full
  HRC-I spikes much above 100 cts/s, the telemetry limit affects the source
  count rate.\label{fig:ltcrv}}
\end{figure}

The full-field lightcurve included significant periods when the count
rate approached the 184 cts s$^{-1}$\ telemetry limit.  The expected
HRC-I background rate for the full field is $\sim 50$\,cts s$^{-1}$
\citep{POG06}.  Despite the brightness of the source, the telemetry
saturation was in fact primarily due to the particle background.
After excluding a 2' radius circle around GX13+1, the average count
rate was $<50$\,cts s$^{-1}$, but with excursions above 100 cts
s$^{-1}$\ where telemetry saturation would affect the data.  To
eliminate this problem only time periods where the total counts in the
field ({\it i.e}, $> 2'$\ from GX13+1) were $<45$\,cts s$^{-1}$\ were
included.  Although this reduced the total good time to 3.58 ksec,
$\sim 200,000$\ counts were detected within $2'$ of GX13+1 for a
source count rate of 54.8 cts s$^{-1}$.  All of these effects can be
seen in Figure~\ref{fig:ltcrv}, which shows the total HRC-I count
rate, the HRC-I count rate within $2'$\ of GX13+1, and the RXTE PCA
lightcurve (between 2-9 keV) during the observation.  Within $2''$\ of
GX13+1 the count rate during these 3.58 ksec was 42.7 cts s$^{-1}$.
According to \S4.2.3.1 of the \chandra Proposer's Observatory Guide,
the encircled energy within $2''$\ is $\approx 90$\%, rising to
$\approx 95\%$\ within $10''$.  Based only on these values, it appears
that $\sim 78$\% of the total counts are ``on-axis'' while 22\% are
scattered by a combination of interstellar dust and the \chandra
mirrors.

\begin{figure*}
\includegraphics[totalheight=2.8in]{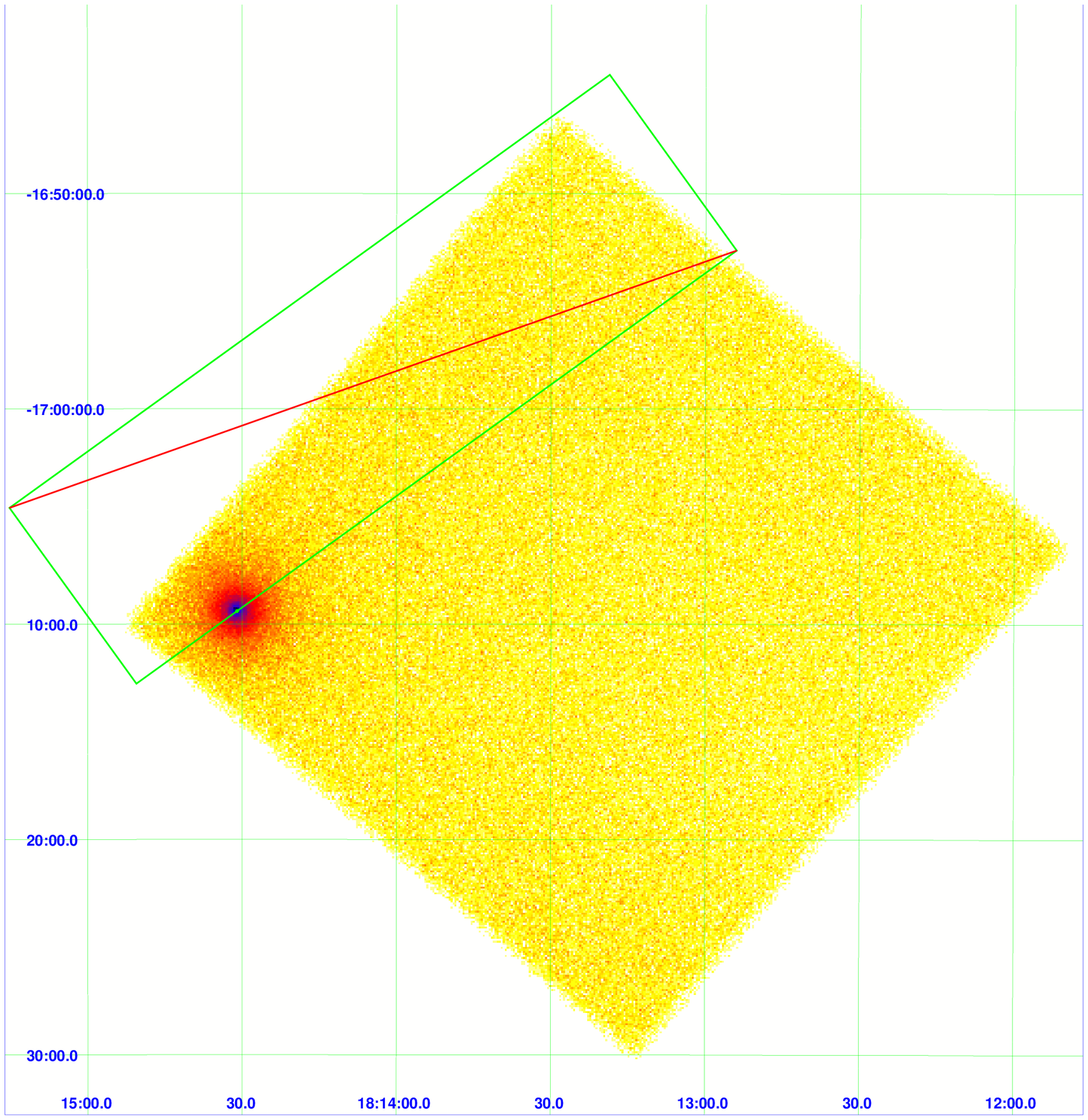}
\hfill
\includegraphics[totalheight=2.8in]{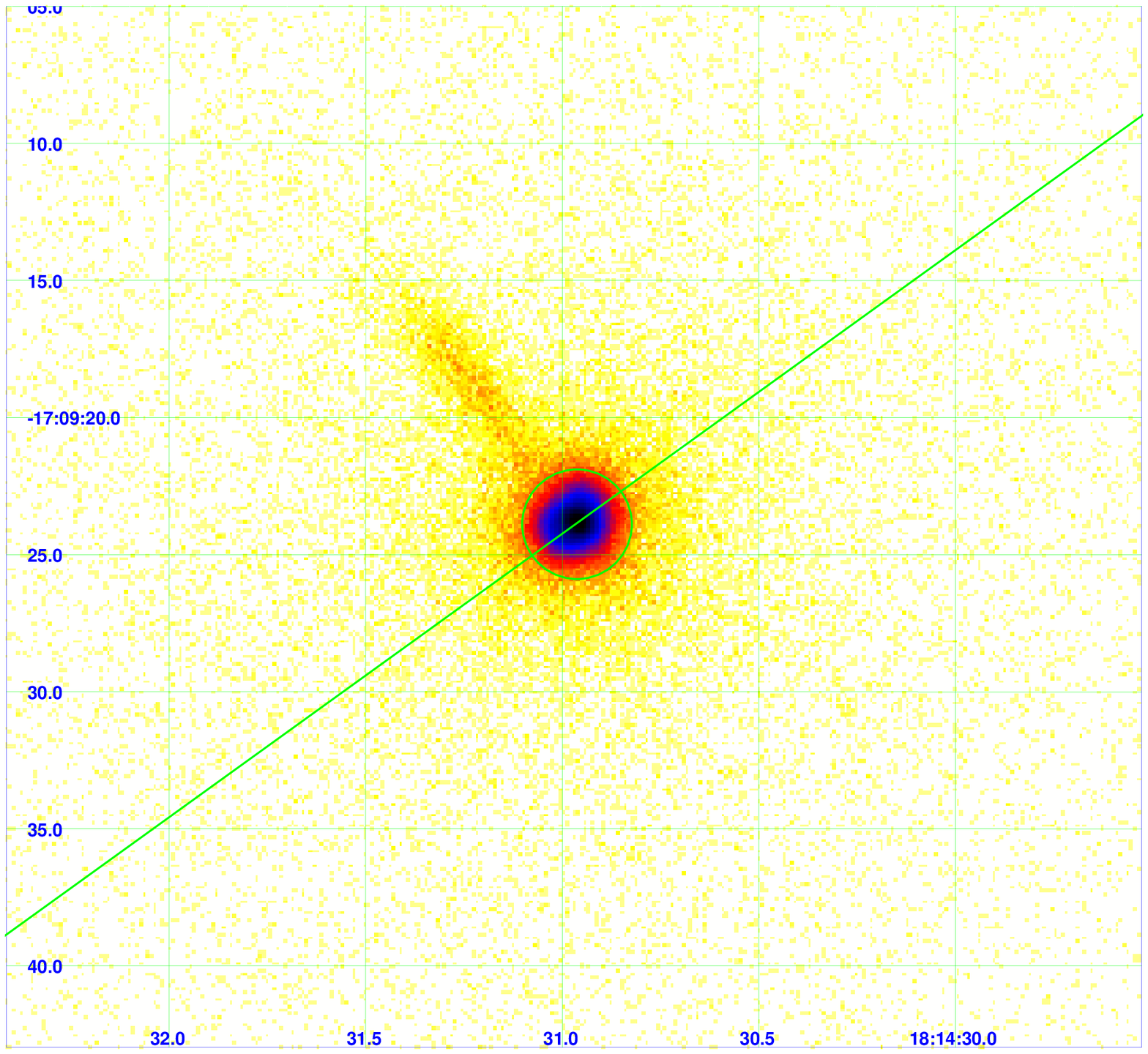}
\caption{[Left] The full HRC-I observation of GX13+1, with the
  excluded rectangular region marked. [Right] Expanded image of GX13+1
  on the HRC-I, showing detector ``jet'' and excluded region.  The 2''
  radius circle shows the near-source region excluded due to likely
  detector non-linearity.\label{fig:image}}
\end{figure*}

The extremely high flux from the source combined with the desire to
get the highest possible spatial resolution required an unusual
instrument configuration.  In collaboration with the CXC Operations
team, the HRC-I detector was positioned so that the source would
appear in one corner of the HRC-I, while still being on-axis to the
HRMA.  This offset retained \chandra's spatial resolution but ensured
the source was far away from the normal aimpoint.
Figure~\ref{fig:image}[Left] shows the full field of the HRC-I, with
GX13+1 at one corner.

In Figure~\ref{fig:image}[Right], a ``jet''extending to the NE and
containing $\sim 1000$\,counts can be seen.  This jet is a well-known
detector artifact \citep{Murray00} which is normally removed by the
standard processing to a level of $<0.1\%$\ of the total source flux
\citep{MurrayGhost}.  In the case of GX13+1, this jet is $\sim 0.5\%$\
of the apparent source count rate.  The most likely cause is the high
source count rate interfering with the the on-board electronic event
processing (Dr. Michael Juda, private communication).  Although the
jet could be eliminated with aggressive filtering, this would also
invalidate the standard calibration.  It was therefore decided to
simply ignore all events from the ``jet''-side of the source, as shown
by the box region in Figure~\ref{fig:image}[Left].

\subsection{Extracting the Spectrum and Flux}

\begin{figure*}
\includegraphics[totalheight=2.2in]{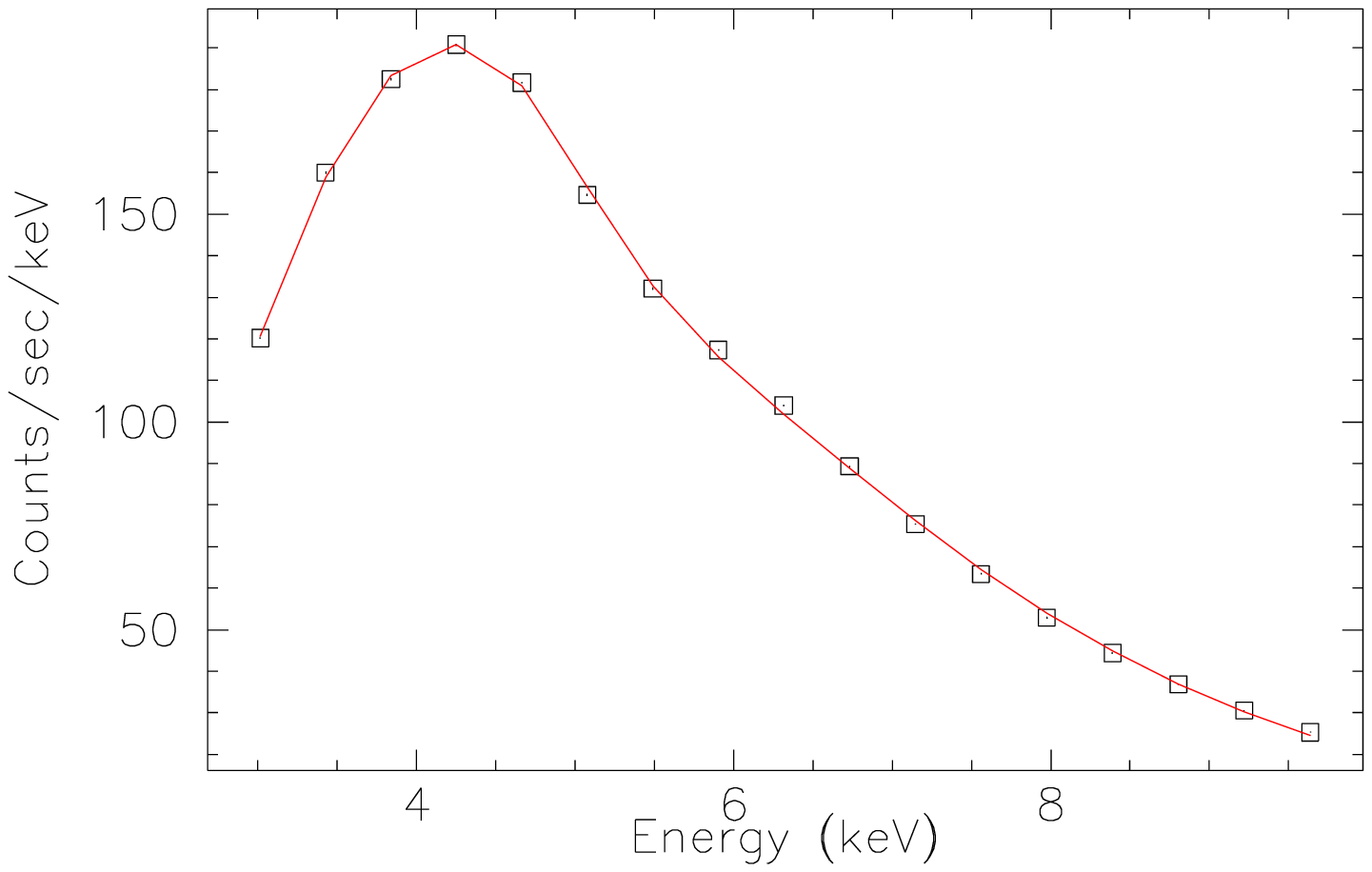}
\includegraphics[totalheight=2.2in]{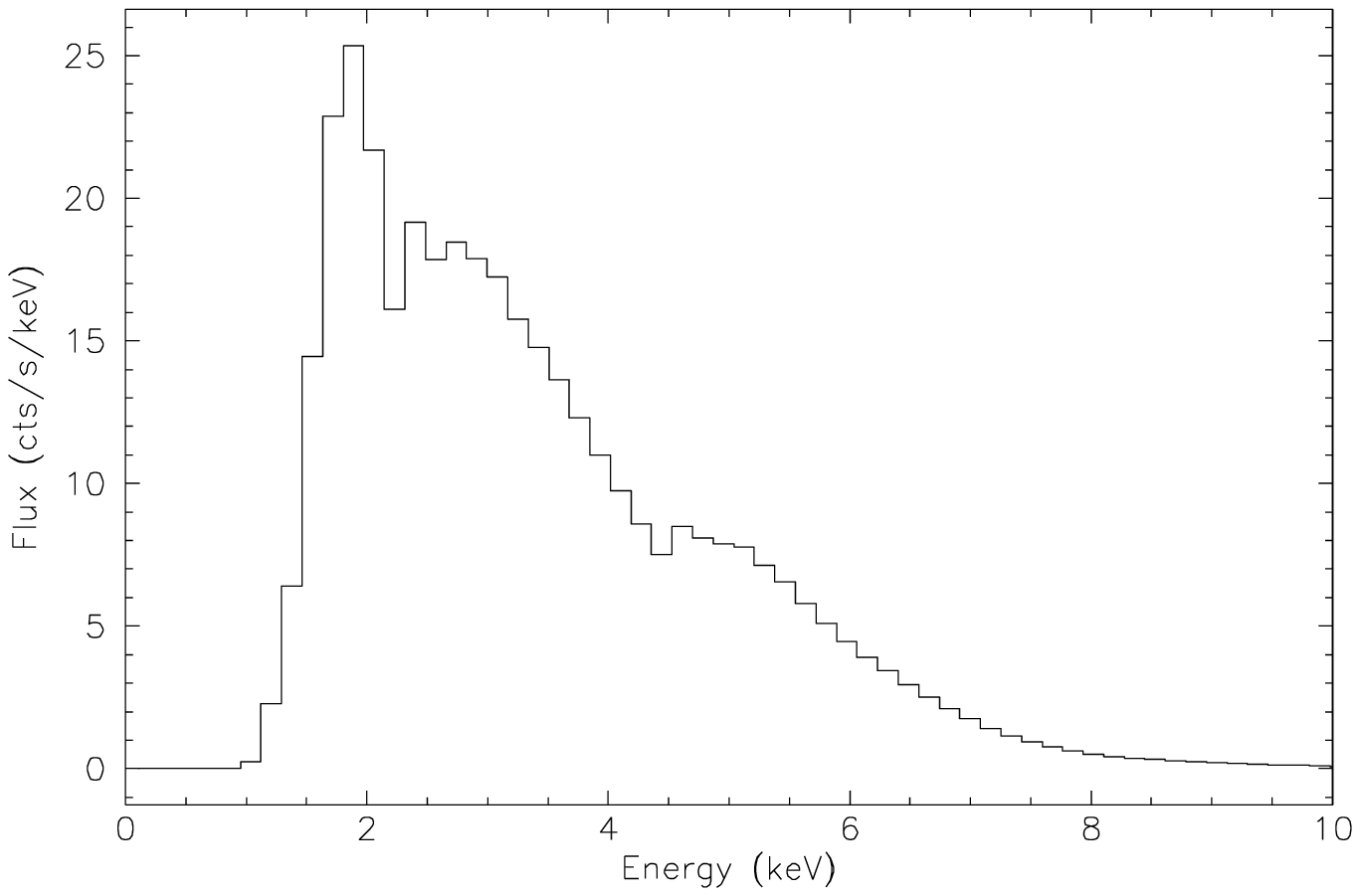}
\caption{[Left] The RXTE PCA spectrum of GX13+1, fit with an absorbed
  multicolor disk plus blackbody model. [Right] The best-fit model
  folded through the HRC-I response.  The integrated count rate
  predicted is 64.6 cts/s.  \label{fig:xtespec}}
\end{figure*}

The surface brightness of the X-ray halo must be normalized by the
source flux to make absolute measurements of the dust column density.
GX13+1 is observed almost constantly since it is a RXTE All-Sky
Monitor (ASM) source with a average rate of 20-30 cts/s.  However, as
the RXTE ASM has little spectral sensitivity and the HRC-I has no
effective energy resolution, simultaneous RXTE PCA observations of
GX13+1 were taken to obtain a useful spectrum of the source.  Although
the RXTE PCA itself has only moderate resolution and little
sensitivity below 2 keV, GX13+1's spectrum is dominated by 2-4 keV
photons (SES02).  The PCA spectrum is shown in
Figure~\ref{fig:xtespec}[Left] as fit with a simple model consisting
of an absorbed multi-color disk model plus a blackbody, following
\citet{Ueda04}.  The column density was fixed at the value found by
\citet{Ueda04} from the Chandra HETG, $_{\rm H} =
3.2\times10^{22}$\,cm$^{-2}$, since this result is far more accurate
than one obtained from the PCA.  The best-fit inner temperature of the
multicolor disk and the blackbody were 1.73 keV and 3.52 keV, with
absorbed 1-10 keV fluxes of $7.43\times10^{-9}$\,erg
cm$^{-2}$s$^{-1}$\ and $2.30\times10^{-10}$\,erg cm$^{-2}$s$^{-1}$,
respectively.  Despite the large reduced $\chi^2_{\nu}$ ($>50$, driven
by systematic errors), this model is an adequate fit for this work
since only the total flux and the approximate spectral shape are
needed to calculate the predicted response of the \chandra HRC-I.
Nonetheless, it is important to note that fluxes measured with the
RXTE PCA are systematically high by 10-15\% in the 2-10 keV range,
compared to other X-ray observatories \citep{Jahoda05}.
%The measured flux from the HRC-I was therefore reduced by 15\% (from
%the RXTE value) to bring the count rates predicted by PIMMS in line
%with the observed values and to account for the discrepancy in the
%RXTE PCA calibration.

To check the expected count rate, we folded this spectrum through the
HRC effective area file for on-axis Cycle 7 data ({\tt
hrciD2005-11-30pimmsN0008.fits}), as shown in
Figure~\ref{fig:xtespec}[Right].  The total predicted source count
rate in the HRC-I is 64.6 cts s$^{-1}$, $\sim 18$\% larger than the
observed HRC-I count rate of 54.8 cts s$^{-1}$\ within 2' of GX13+1.
% see makeplot_ueda.sl for details 
The discrepancy is primarily due to the overestimation from the RXTE
PCA calibration, with an additional complication due to spatial
variation in the response of the HRC-I that reduces the effective area
of the detector corners relative to the center \citep{Donnelly03} 

The \chandra PSF, measured as a ratio of the surface brightness to the
source flux, is the background for this observation.  The RXTE PCA, a
non-imaging detector, includes both the direct source flux and the
scattered halo photons, which must be removed to avoid
double-counting.  However, as the goal is to measure the scattered halo
fraction itself, this problem is recursive.  I addressed this by
assuming a column density of $3.2\times10^{22}$\,cm$^{-2}$\ and
calculating the total scattered fraction for the MRN77, WD01, and
ZDA04 BARE-GR-B models, weighted by the HRC-I response.  The resulting
halo fraction ranged from 13-26\%.  This predicted halo strength is
consistent with the result that 22\% of the total source counts are
between $2''-120''$.  Therefore, for purposes of calculating the
background PSF, the RXTE PCA flux was reduced by 20\% to exclude the
halo contribution, with a 7\% systematic error.  This reduction is in
addition to the 15\% reduction described above.  The 7\% error is
likely not the dominant term in the systematic error, however.  The
observation of a very bright source in one corner of the HRC-I
detector is at the extreme edge of the available calibration, and so
careful consideration of all uncertainties will be required.

Another concern regarding this observation was that a significant
short-term change in the source flux, on the order of 12-24 hours,
would also affect the halo in a time-delayed manner \citep[\protect{\it
e.g.}][]{Vaughan04}. At smaller angles the delay could be even longer.
The RXTE ASM data was checked for a 10 day period before the
observation, but no strong or significant variation was seen.
Although Type I X-ray bursts have been seen from GX13+1 which show
$3-4\times$\ the normal flux, they only last $\sim 15$\,seconds
\citep{Matsuba95}.  In this case, no bursts were seen in either the
HRC-I or PCA lightcurves, and indeed the halo observation would not be
sensitive to such a small variation.

\subsection{Point-spread function\label{ss:psf}}

An accurate measurement of the Chandra HRC-I point-spread-function
(PSF) between 2-100'' from the source is crucial to this observation.
An accurate raytrace model (ChaRT
\footnote{http://asc.harvard.edu/soft/ChaRT/cgi-bin/www-saosac.cgi}),
of the Chandra HRMA has been calibrated for near-source ($< 2''$)
photons
%\footnote{http://asc.harvard.edu/cal/docs/cal_present_status.html\#psf}
, SES02 showed that at large scattering angles this model
significantly underestimates the PSF, leading to substantial problems
in the analysis.  Therefore, SES02 relied upon an ACIS-I observation
of Her X-1 as a PSF calibrator, but this source is affected by pileup
within $\sim 10$'' and therefore cannot be used in the 2-10'' range.

\begin{figure}
\includegraphics[totalheight=2.3in]{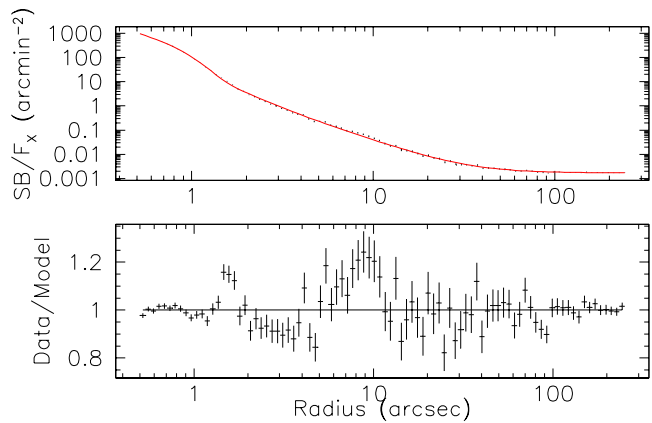}
\caption{[Top] The radial profile of 3C273's surface brightness,
  divided by the source flux, fit with the sum of a Gaussian plus two
  power laws and a constant. [Bottom] The ratio of the data and model,
  showing small excursions at $1.5''$\ and $10''$, but generally good
  agreement. \label{fig:PSF_3C273}}
\end{figure}

As shown in Figure~\ref{fig:xtespec}[Right], the spectrum of GX13+1
peaks at $\sim 2$\,keV.  The HRC-I's lack of spectral response means
that spectral differences between any calibration source and GX13+1
will lead to additional complications.  The best possible calibration
source would be a bright, hard, and lightly-absorbed X-ray source
observed on-axis with the HRC-I.  The X-ray binary LMC X-1 matches
these requirements reasonably well, and two \chandra observations
(ObsID 1200, 1201) of the source have been done.  However, they were
both done early in the mission (August 1999) before the HRMA final
focus was set and are thus unsuitable.  Since then, the brightest hard
X-ray source with little absorption and a known (albeit variable) flux
to be observed with the HRC-I is 3C273 (ObsID 461 on Jan 22, 2000).
Figure~\ref{fig:PSF_3C273} shows 3C273's surface brightness, divided
by its source flux, as observed with the \chandra HRC-I (excluding the
well-known jet region).  The spectrum was taken from a \chandra HETG
observation done twelve days earlier (ObsID 459) which is well-fit by
an absorbed power-law with $\Gamma = 1.67\pm0.01$\ and $F_X(2-10{\rm
keV}) = (1.08\pm0.03)\times10^{-10}$\,ergs cm$^{-2}$s$^{-1}$.  The
absorption column was fixed at the Galactic value, N$_{\rm H} =
1.8\times10^{20}$\,cm$^{-2}$.  The predicted HRC-I count rate (based
on the CXC PIMMS tool) for this spectrum is 8.7 cts/s, while the
actual source count rate was 26\% higher at 11 cts/s.  As the source
is variable, this was taken as showing little change and the flux was
simply assumed to have increased by 26\% during the HRC-I observation.
A significant but undetected change in the spectral shape could also
cause this change in observed flux and might affect our results.
However, BeppoSAX observations of 3C273 over a period of 10 days
showed only small changes in the 2-10 keV spectral shape, with
$\Gamma$\ values ranging from $1.56\pm0.02$ to $1.64\pm0.02$\ while
the flux varied by 15\% \cite{Haardt98}.  The potential systematic
error caused by uncertainty in the true spectrum of 3C273 during its
observation with the HRC-I is thus smaller than the error due to
unavoidable differences between the spectra of 3C273 and GX13+1.

The core of the PSF of 3C273 was fit with a Gaussian term centered at
0 with FWHM of $1.007_{-0.005}^{+0.004}$''\ and amplitude
$1676\pm42$\,arcmin$^{-2}$.  In addition, the best-fit model included
two power-law terms with $\Gamma_1 = 4.06\pm0.05, \Gamma_2 =
2.40_{-0.02}^{+0.01}$\ and amplitudes $A_1 =
(3.4_{-0.9}^{+1.3})\times10^{-7}, A_2 =
(2.12_{-0.09}^{+0.13})\times10^{-4}$\ arcmin$^{-2}$\ at 100''.  The
particle and sky background was fit with a constant,
$(2.60\pm0.02)\times10^{-3}$\,arcmin$^{-2}$.  As
Figure~\ref{fig:PSF_3C273} shows the fit is quite good over a large
range of surface brightnesses, with the somewhat large reduced
$\chi^2_{\nu} = 2.6$\ likely due to the extreme precision of the
measurement compared to the relatively simple model.

\section{Results}

The HRC-I observations are most useful between 2-100'', since beyond
that radius the ACIS-I data can measure the energy-resolved X-ray
halo.  Therefore, the ACIS-I data were reprocessed (with CIAO 3.3) and
reanalyzed following the approach described in SES02 except as noted
below.  Both the HRC-I and ACIS-I results were used in the final
analysis.  We note that in reprocessing the ACIS data, the source flux
measurement, done via the CCD transfer ``streak'', was redone with a
better calibration and improved handling of the background subtraction
which resulted in an overall $\sim 15$\% decrease in the measured
source flux.  The calibration changes include a spatially-varying
modification of order $\pm5\%$\ in the quantum efficiency uniformity
in CALDB 2.28, and an energy-dependent increase of up to 16\% in the
overall effective area which was added in CALDB v3.2.1.

The data were fit using the CIAO fitting engine {\sl Sherpa}\ using
scattering models based on the exact Rayleigh-Gans (RG) approximation
\citep{SD98}.  This model assumes the grains are spherical but uses
the energy-dependent optical constants rather than the Drude
approximation when calculating the scattering efficiency.
\citet{SD98} noted that the full Mie treatment is necessary for X-rays
$< 2$\,keV, since the RG overestimates the total scattering at low
energies.  The HRC-I is sensitive to X-rays between 0.08-10 keV, with
peak efficiency between 0.7-2.0 keV.  In all cases the halo model for
the HRC-I was calculated using an average value weighted by the
spectrum of GX13+1 and efficiency of the HRC-I.  However, since
GX13+1's spectrum as observed by the HRC-I (see
Figure~\ref{fig:xtespec}[Right]) is dominated by photons with $E >
2$\,keV, the use of the simpler RG treatment is justified.

The initial analysis assumed the dust was smoothly distributed along
the line of sight.  Unlike SES02, where the predicted PSF was
subtracted from the data, here the PSF was incorporated into the
fitting directly to allow for an explicit inclusion of uncertainty in
the PSF.  Fits to the ACIS-I and HRC-I data included a constant factor
that allowed for calibration uncertainty in the overall PSF, caused
primarily by systematic errors in the source flux.  For both the ACIS-I
and HRC-I data this multiplier was allowed to vary by up to 10\%.  In
many cases the fit pushed the multiplier to an extremum of the range,
showing that systematic uncertainties remain in the data, although it
is not clear what component dominates them.  In SES02, systematic
errors in the PSF manifested as energy-dependent column density fits,
since to first order an error in the PSF could be adjusted by changing
the overall halo scattering.  As the total halo intensity is inversely
proportional to energy, this effect is often a linear dependence of
the best-fit N$_{\rm H}$\ on energy.  To check for this, I allowed the
value of N$_{\rm H}$\ to vary independently in the HRC-I and each energy
band of the ACIS-I data.  I used the F-test to determine that in only
one case (ZDA04 BARE-GR-B) were the best-fit ACIS-I N$_{\rm H}$\ values
better described by a linear energy-dependent model than by a constant
value.  Even in this case, the F-test significance was only 3.5\%, a
negligible value given the number of different models tried.  It seems
unlikely, therefore, that there is a significant error in the relative
power in the dust-scattered (halo) and mirror-scattered (PSF) photons.

Table~\ref{tab:smofits}\ shows the best-fit N$_{\rm H}$\ results for
the HRC-I and the ``average'' ACIS-I value fit and the total
$\chi^2_{\nu}$\ assuming smoothly-distributed dust along the line of
sight for the MRN77, WD01, and the 15 ZDA04 models.  These can be
compared to the value of $3.2\times10^{22}$\,cm$^{-2}$\ found by
\citet{Ueda04}.  The ACIS-I column densities are $\sim 20\%$\ larger
than the HRC-I values, although the models with the lowest
$\chi^2_{\nu}$\ values tend to have the best agreement.  The most
likely cause of this discrepancy is cumulative errors in the source
flux measurements combined with calibration differences between the
ACIS-I and HRC-I detectors.
% need to rerun smooth zda7 

\begin{table}[t]
\caption{Smooth Dust Model Parameters\label{tab:smofits}}
\begin{tabular}{llll}
\hline \hline
Model     &N$_{\rm H}$(HRC)& N$_{\rm H}$(ACIS) & $\chi^2_{\nu}$ \\
          &$10^{22}$\,cm$^{-2}$&$10^{22}$\,cm$^{-2}$&  \\ \hline
MRN77     &$2.4\pm0.2$  &$2.85\pm0.05$& 2.0 \\
WD01      &$1.51\pm0.02$&$2.0\pm0.2 $ & 3.6 \\
BARE-GR-S &$2.7\pm0.2$  &$2.9\pm0.1 $ & 1.9 \\
BARE-GR-FG&$2.6\pm0.2$  &$3.00\pm0.04$& 1.9 \\
BARE-GR-B &$3.6\pm0.3$  &$3.4\pm0.4$  & 2.8 \\
BARE-AC-S &$2.5\pm0.2$  &$3.0\pm0.1$  & 2.1 \\
BARE-AC-FG&$2.5\pm0.1$  &$3.0\pm0.2$  & 2.2 \\
BARE-AC-B &$3.3\pm0.3$  &$3.60\pm0.05$& 2.2 \\
COMP-GR-S &$2.02\pm0.02$&$2.9\pm0.4$  & 4.9 \\
COMP-GR-FG&$2.23\pm0.04$&$3.0\pm0.3$  & 3.5 \\
COMP-GR-B &$3.0\pm0.2$  &$3.71\pm0.09$& 1.9 \\
COMP-AC-S &$2.41\pm0.02$&$3.7\pm0.6 $ & 7.0 \\
COMP-AC-FG&$2.67\pm0.02$&$3.8\pm0.5 $ & 5.0 \\
COMP-AC-B &$4.24\pm0.04$&$6.7\pm0.9$  & 5.7  \\
COMP-NC-S &$11.1\pm0.6$ &$13.5\pm0.9$ & 2.4 \\
COMP-NC-FG&$2.81\pm0.02$&$4.6\pm0.8$  & 8.6 \\
COMP-NC-B &$3.39\pm0.03$&$5.9\pm1.1$  & 11.7 \\ \hline
\end{tabular}
\end{table}

\begin{figure*}
\includegraphics[totalheight=2.3in]{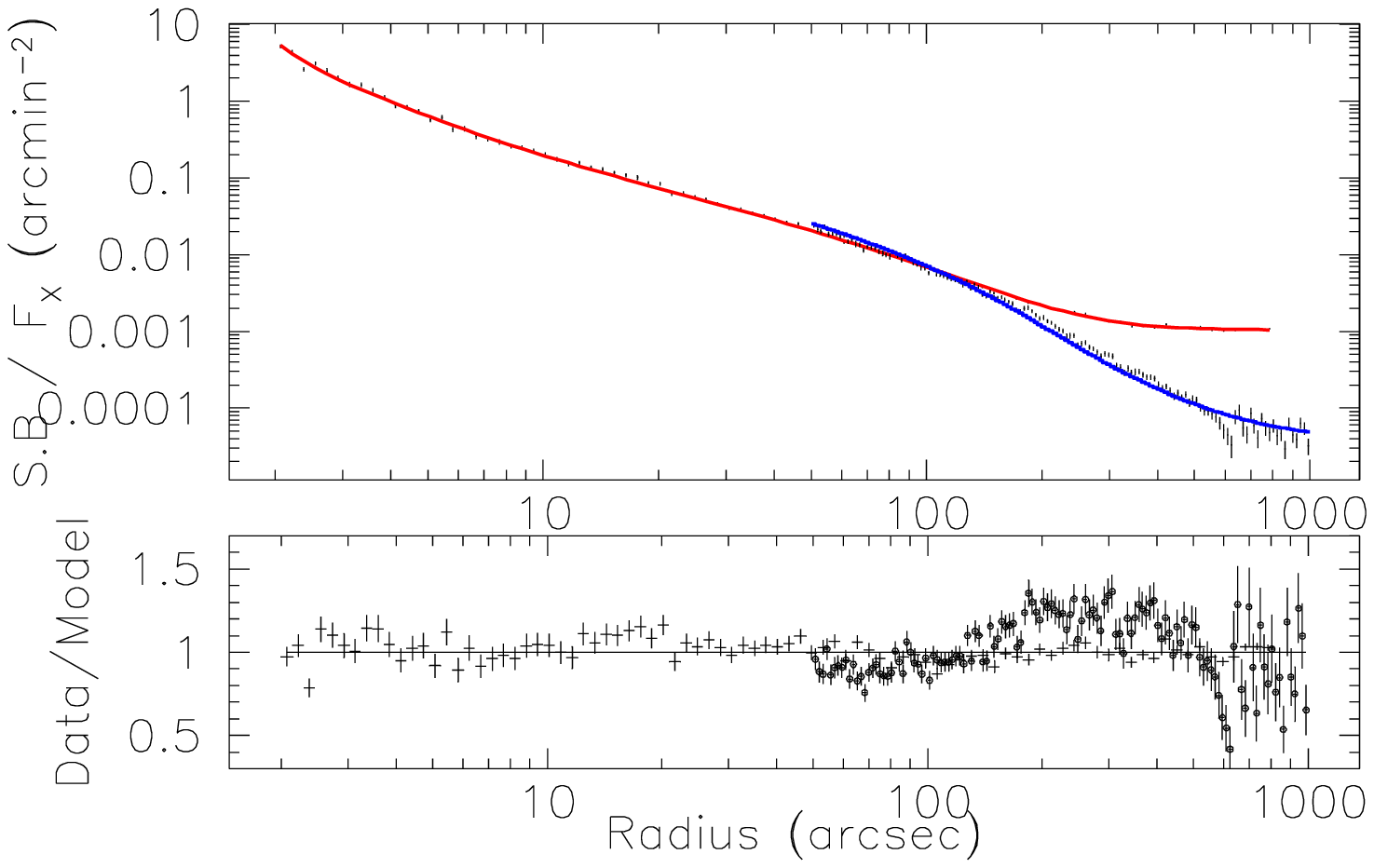}
\includegraphics[totalheight=2.3in]{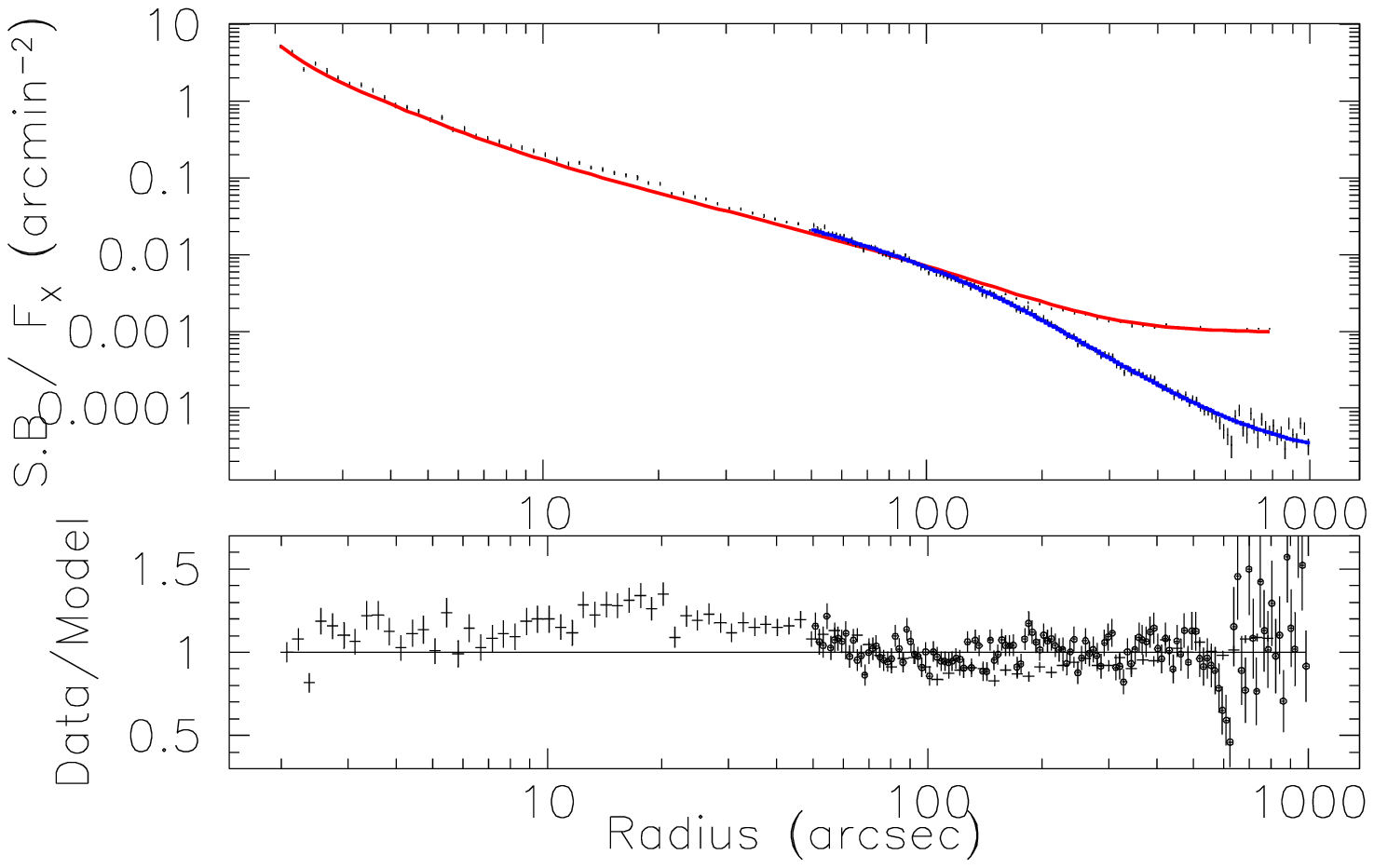}
\caption{[Left] The X-ray halo from smoothly-distributed WD01-type
  grains fit to the radial profile of GX13+1's surface brightness
  divided by the source flux.  The HRC-I observations are fit with a
  red line; the ACIS-I data (at 2.5 keV) with a blue line.  The ratio
  of the HRC-I data to the model is shown below with crosses and the
  ACIS-I data with circles.  [Right] Same, for the ZDA04 BARE-GR-S
  model.
  \label{fig:Smooth}}
\end{figure*}

Figure~\ref{fig:Smooth} shows the profile of the HRC-I data along with
the ACIS-I data at $2.5\pm0.1$\,keV, near the median energy of the
spectrum as observed by the HRC-I.  This figure shows the level of
agreement between the HRC-I and ACIS-I data agree with each other in
the overlap region ($50-100''$), as well as the large ($> 20\times$)
difference in the HRC-I and ACIS-I backgrounds in the $500-1000''$\
region.  The radial profile shown in Figure~\ref{fig:Smooth}\ is fit
assuming the line of sight (LOS) dust is ``smoothly-distributed'' and
has a composition and size distribution described by WD01 [Left], and
the ZDA04 BARE-GR-S [Right] models.  Although both models agree with
the overall shape of the radial profile, the ratio plots show that the
WD01 model underestimates the ACIS-I data in the $300-500''$\ range,
while the BARE-GR-S model underestimates the HRC-I data in the
$10-50''$\ range.

The smoothly distributed dust fits show small discrepancies that might
be indications of dusty molecular clouds along the line of sight.
These would appear as ``bumps'' in the profile whose position and
strength depends upon the relative distance to the cloud and its
column density.  I therefore refit the HRC-I and ACIS-I data using a
two-component model that included a smoothly-distributed component
plus a thin cloud with variable position and column density; the cloud
is treated as a sheet with negligible thickness.  Models using only a
single thin cloud with no smooth component were also considered; these
gave generally poor fits independent of the dust model used, and so
were abandoned.  This is not unexpected since (a) SES02 was unable to
fit a single cloud using only the ACIS-I data and (b) GX13+1 is
reasonably near the Galactic center ($(l,b) = (13.5^{\circ},
0.1^{\circ})$, $D = 7\pm1$\,kpc; \citet{Bandyopadhyay99}) where a
sightline dominated by a single cloud would be unusual.

To reduce fit time, the column density of both the smooth component
and a cloud was fixed to be the same for all datasets, as was the
position of the cloud along the line of sight.  While more realistic
than allowing the cloud column density to vary as a function of X-ray
energy, this has the effect of magnifying residual systematic errors.
As noted previously, the halo strength diminishes with energy while the
relative PSF strength increases which can create a trend in the
best-fit column density as a function of energy.  However, since in
only one case out of fourteen was such a trend seen previously, it
seems unlikely that the systematic errors are driving the resulting
best-fit parameters in the smooth plus cloud model.

The best-fit parameters for each model are shown in
Table~\ref{tab:clfits}.  None of the fits are formally acceptable
($\chi^2_{\nu}$\ ranges from 2.1 to 9.9), although some are clearly
better than others.  In cases where the best-fit position is 0, the
1$\sigma$\ upper limit is shown.

\begin{table}[t]
\caption{Smooth Plus Cloud Model Parameters\label{tab:clfits}}
\begin{tabular}{lllll}
\hline \hline
Model     &N$_{\rm H}$(smooth) &N$_{\rm H}$(cloud)  & Relative & $\chi^2_{\nu}$ \\
          &$10^{22}$\,cm$^{-2}$&$10^{22}$\,cm$^{-2}$& Position &  \\ \hline
MRN77     &$2.60_{-0.05}^{+0.08}$&$0.69_{-0.03}^{+0.06}$&$<0.003$&2.4 \\
WD01      &$1.38\pm0.02$&$0.46\pm0.01$&$<0.001$ &3.5 \\
BARE-GR-S &$3.02\pm0.01$&$0.15\pm0.01$&$0.89\pm0.01$&2.1 \\ %4
BARE-GR-FG&$2.95\pm0.01$&$0.11\pm0.01$&$0.91\pm0.01$&2.2 \\ %5
BARE-GR-B &$3.51\pm0.01$&$0.57\pm0.01$&$0.82\pm0.01$& 2.4\\ %6  
BARE-AC-S &$2.6\pm0.1$  &$0.21\pm0.06$&$<0.001$ &2.4 \\	    %7 
BARE-AC-FG&$2.6\pm0.1$  &$0.23\pm0.06$&$<0.001$ & 2.4\\	    %8 
BARE-AC-B &$3.4\pm0.1$  &$0.35\pm0.14$ &$0.85\pm0.01$& 2.1 \\	 %9 
COMP-GR-S &$1.6\pm0.1$  &$0.82_{-0.09}^{+0.07}$&$<0.0003$& 2.9\\ %10
COMP-GR-FG&$2.0\pm0.1$  &$0.62_{-0.08}^{+0.06}$&$<0.004$&2.6 \\ %11
COMP-GR-B &$3.2\pm0.1$  &$0.23\pm0.05$&$<0.01$&2.2 \\       %12
COMP-AC-S &$1.6_{-0.1}^{+0.2}$&$1.3\pm0.1$&$<0.0002$ &3.7 \\%13     
COMP-AC-FG&$2.2_{-0.1}^{+0.2}$&$1.0\pm0.1$&$<0.0003$&3.3 \\ %14     
COMP-AC-B &$2.9\pm0.2$  &$2.3\pm0.1$&$<0.0004$ &3.6 \\      %15
COMP-NC-S &$11.4_{-0.3}^{+0.6}$&$1.3\pm0.3$&$<0.0008$& 2.5\\%16
COMP-NC-FG&$1.6_{-0.2}^{+0.3}$&$1.9\pm0.2$&$<0.0001$ &4.3 \\%17
COMP-NC-B &$1.6_{-0.5}^{+0.9}$&$2.5_{-0.5}^{+0.4}$&$<0.00001$ &6.5 \\ \hline
\end{tabular}
\end{table}

\section{Discussion}

\citet{SES02} analyzed the ACIS-I observations of GX13+1's dust
scattered halo and found that the dust size distribution does not
extend to very large ($> 1\mu$m) grains and that grains do not have a
large vacuum fraction.  However, the ACIS-I data could not distinguish
between the MRN77 and WD01 models, as the two distributions lead to
similar scattering profiles at large angles.  Similarly, the data left
open the possibility that there might be a substantial population of
grains near the source \citep{Draine03,Xiang05}.  These would create a
near-source scattered halo that was obscured by pileup in the ACIS-I
detector.  The primary goal of the HRC-I observation was to remove
these uncertainties by measuring the halo near the source.  This would
determine which dust model best fit the data, as well as detecting (or
put limits upon) variation in the dust distribution along the line of
sight.

The fit results shown in Tables~\ref{tab:smofits} and \ref{tab:clfits}
contain a few surprises.  Just as in \citet{SES02}, the WD01 model had
the smallest column density of any of the models when fit with either
smoothly distributed dust or after adding a cloud.  However, the
overall result was a significantly worse fit than found with either
the MRN77 or many of the ZDA04 models.  As Figure~\ref{fig:Smooth}
shows, the smooth WD01 model fits the HRC data well, but
underestimates the halo measured by ACIS between $150''-400''$, while
the ZDA04 BARE-GR-S model underestimates the halo measured by the HRC
between $10-50''$.  Examining the other models show that these two
cases are representative.  Adding a single dust cloud to the model
results in a solution with a cloud 70-90\% of the distance to the
source if the pure smooth model underestimates the halo around $30''$.
Conversely, adding a cloud component to smooth dust models that
underestimate the halo around $300''$\ tend to put the cloud near the
Sun.

\begin{figure*}
\includegraphics[totalheight=2.2in]{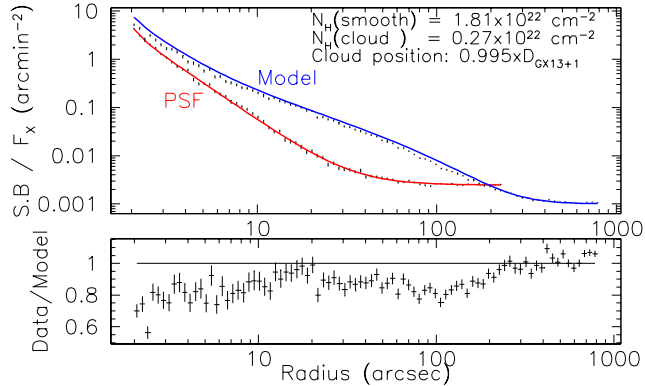}
\caption{[Top] Radial profile of GX13+1 from the HRC-I (upper points) with
  PSF calibration data from 3C273 (lower points).  A close
  approximation to the \citet{Xiang05} model for GX13+1 using the WD01
  model is shown in blue, with the PSF fit to the 3C273 data is shown
  in red. [Bottom] Ratio of GX13+1 data to \citet{Xiang05} model.
\label{fig:xiangwd01}} 
\end{figure*}

Although uncertainties remain due to calibration issues, the overall
quality of the fits shown in Table~\ref{tab:smofits} and
Figure~\ref{fig:Smooth} do not support the proposition that a
significant cloud of dust is present near GX13+1.  In particular,
although it is not identical to the \citet{Xiang05} model,
Figure~\ref{fig:xiangwd01} shows the result from a model similar to
their fit to the zero-order HETG for GX13+1 using WD01-type dust.
This approximation to their model puts a cloud with N$_{\rm H} =
2.7\times10^{21}$\,cm$^{-2}$\ at a position 99.5\% of the distance to
GX13+1, along with a smooth distribution with N$_{\rm H} =
1.81\times10^{22}$\,cm$^{-2}$.  In either their model or my
approximation, the cloud near the source dominates the variation in
the dust distribution.  Figure~\ref{fig:xiangwd01} shows that although
the HRC-I data contain an obvious halo, this model is not a good fit
and, in fact, removing the cloud improves the fit.  Although their
model fit can be improved somewhat with a small vertical offset
(corresponding to different relative flux calibrations of $\sim
12\%$), the model still predicts 10-20\% more halo for $\theta <
10''$\ than is observed.  Table~\ref{tab:clfits} confirms this
conclusion, finding (in the case of WD01-type dust) a cloud near the
Sun, not the source, with N$_{\rm H} \sim 5\times10^{21}$\,cm$^{-2}$.

ZDA04 described in detail how constraining dust models requires
combining multiwavelength data from the IR to X-rays while
simultaneously considering the metal abundances in the grains.  Due to
the nature of optical/UV extinction and X-ray scattering, few sources
show strong signatures of dust in all of these wavebands
\citep{ValencicSmith07}.  Nonetheless, it is possible to constrain the
allowed dust models by comparing the column density predicted by the
models to that measured using other techniques.  In the case of
GX13+1, measurements of the column density range from
$2.5-4.0\times10^{22}$\,cm$^{-2}$\ \citep{CN92}.  Optical measurements
provide only an upper limit of $2.9\times10^{22}$\,cm$^{-2}$\ based on
plausible but unconfirmed assumptions about the source
spectrum\citep{Garcia92}.  The total \ion{H}{1} column density through
the Galaxy at the position of GX13+1 is $1.8\times10^{22}$\,cm$^{-2}$
\citep{DL90}, but this is misses the contribution from molecular
H$_2$\ that is likely to be substantial in the Galactic plane.  The
HETG observation of GX13+1 agrees (weakly) with these results
\citep{Ueda04}, although it does not strongly limit it.  Only Mg can
be directly measured ($N_{Mg} = 1.84^{+0.91}_{-0.49} \times
10^{18}$\,cm$^{-2}$), equivalent to \NH $=
4.8^{+2.4}_{-1.3}\times10^{22}$\,cm$^{-2}$\ assuming solar abundances.
The $2\sigma$\ upper limits for Si and S are equivalent to $\NH <
4\times10^{22}$\,cm$^{-2}$.  Of course, LMXBs have shown significant
variable internal absorption \citep{HG83} in X-rays, so this spectral
measurement sets at best an upper limit to the actual interstellar
component that is responsible for the halo.

Despite these difficulties in independently measuring the total LOS
dust column density, we can reasonably justify excluding the value of
N$_{\rm H} = (1.11\pm0.06)\times10^{23}$\,cm$^{-2}$\ found in
Table~\ref{tab:smofits} for the ZDA04 COMP-NC-S model fit to the HRC-I
data.  However, this was {\it only} model from ZDA04 that used
composite grains without bare carbon grains that had a plausible value
of $\chi^2_{\nu}$.  Although more data are needed, this class of
models, along with the group of ``composite grains with bare amorphous
carbon'' models are clearly suspect since they do not generate an
X-ray halo similar to these observations.  In fact, the only
smoothly-distributed ZDA04 composite grains model that fit with
$\chi^2_{\nu} < 3$\ had graphitic carbon and B star abundances
(COMP-GR-B).  After adding a dust cloud to the model, the COMP-GR-B
model fit with $\chi^2_{\nu} = 2.2$\ while the next best fit
(excluding the unrealistic COMP-NC-S model) was COMP-GR-FG with
$\chi^2_{\nu} = 2.6$, a significantly worse fit.

\section{Conclusions}

The principal results from this analysis are:
\begin{enumerate}
\item Although challenging, HRC-I observations can be used to recover
  the near-source region excluded by pileup in the ACIS-I detector.  The
  lack of energy resolution can be finessed if another measurement of
  the source spectrum is available. 
\item Fitting the source profile and background PSF independently
  improves overall results, since calibration uncertainties in the
  flux from the source and background objects can then be included
  explicitly.
\item There is no strong signature of a dust cloud at the source in
  the radial profile, as suggested by \citet{Xiang05}, although some
  models include a cloud $\sim 90$\% of the distance to the source.
\item In agreement with SES02, the WD01 model underestimates the total
  column density to the source, and again leads to poor fits, although
  not so bad as to be excluded given the calibration uncertainties.  
\item Some models from the ZDA04 paper, if not conclusively excluded,
  are at least implausible.  In general, the ZDA04 models with
  composite grains (excepting the graphitic carbon model with B star
  abundances) gave poor fits, while the bare carbon and silicate grain
  models tended to fit well.
\end{enumerate}

It should be noted that the relatively good fits found using the
simple smoothly-distributed dust model are somewhat surprising, since
X-ray halos probe both the largest grains whose size and composition
are the least constrained from observations in other wavelengths.
Additionally, X-ray halos are primarily observed through
highly-absorbed lines of sight.  These probe dust in dense molecular
clouds that cannot be observed in the optical or UV due to the
extremely large extinction.  Finally, all of these models assume
spherical grains, although recently some calculations have been done
on aspherical grains \citep{DraineAllaf06}\ that show the effects
are small ($<20\%$) for the WD01 model at $\sim 2$\,keV.  Despite
these potential problems, a number of existing grain models agree
quite well with the observations, suggesting grains in dense clouds
(with the exception of the densest regions that take up very little
volume and may be optically-thick to X-rays) are not too dissimilar
from grains in less dense regions.  

\acknowledgments I thank Michael Juda for his substantial assistance
in arranging the observation and understanding the HRC-I calibration.
I would also like to thank Terry Gaetz and Diab Jerius for their
assistance in understanding the \chandra PSF, Lynne Valencic for
helpful discussions about dust, and Jingen Xiang for clarification of
his dust cloud models.  Finally, my appreciation goes out to
Eli Dwek for many helpful discussions and for first bringing the issue
of X-ray scattering in dust to my attention.  This work was supported
by the NASA Chandra observation grant GO5-6144X and by the NASA LTSA
grant NNG04GC80G.

\end{document}